\documentclass[showpacs,amsmath,amssymb]{revtex4}
\usepackage{bm}

\begin{document}

\newcommand{\uu}[1]{\underline{#1}}
\newcommand{\pp}[1]{\phantom{#1}}
\newcommand{\be}{\begin{eqnarray}}
\newcommand{\ee}{\end{eqnarray}}
\newcommand{\ve}{\varepsilon}
\newcommand{\ii}{\iota}
\newcommand{\vs}{\varsigma}
\newcommand{\Tr}{{\,\rm Tr\,}}
\newcommand{\pol}{\textstyle{\frac{1}{2}}}
\newcommand{\lbar}{l_{^{\!\bar{}}}}

\title{Teleportation seen from space-time: On two-spinor aspects of quantum information processing}
\author{Marek Czachor}
\affiliation{
Katedra Fizyki Teoretycznej i Informatyki Kwantowej\\
Politechnika Gda\'nska, 80-952 Gda\'nsk,
Poland\\
and\\
Centrum Leo Apostel, \\
Vrije Universiteit Brussel, 1050 Brussels, Belgium\\
}

\begin{abstract}
Formal similarity between Minkowski tetrads and Bell bases allows to think of metric tensors in terms of quantum teleportation protocols. The role of null tetrads for quantum information processing is different. They define qubits resistant to a special kind of noise that occurs if coding and decoding of quantum information is performed in different reference frames. These examples show that mutual links between quantum information and the 2-spinor calculus may be nontrivial and worthy of further studies. 
\end{abstract}
\maketitle

\section{Introduction}

The 2-spinor abstract index calculus \cite{PR} is a formalism designed for four-dimensional space-times with Lorentzian geometry. The fundamental building block is here a 2-spinor, a two-dimensional complex vector or vector field. The basic operation that leads from 2-spinors to higher-rank objects, including world-vectors and tensors, is the tensor product. Four-dimensional world-vectors, such as position vectors in Minkowski space, are linear combinations of simple tensors composed of pairs of primed and unprimed 2-spinors. Simple tensors play a privileged role and correspond to null world-vectors.

An analogous mathematical structure occurs in quantum information theory. Qubits, similarly to 2-spinors, are two-dimensional complex vectors or vector fields (typically defined on mass hyperboloids in momentum space). Tensor products of qubits lead to binary coding, and non-factorizable linear combinations of simple tensor products of qubits are known as entangled states. Simple tensors involve no entanglement and are of limited use for quantum information processing.

Relations between 2-spinors and qubits involve interesting theoretical subtleties. Not all 2-spinors are qubits, and not all qubits are spinors. Perhaps the most important difference is in transformation properties: Qubits are assumed to transform unitarily whereas 2-spinors may carry non-unitary representations. In curved space-times 2-spinors transform under local versions of SL(2,C). Relativistic qubits transform in momentum space under local versions of SU(2) even in Minkowski space --- this is the price for unitarity of representations of relativistic symmetry groups. Non-unitary finite-dimensional representations in position space are related to infinite-dimensional unitary representations in momentum space by a kind of duality, similar to that between passive transformations of coordinates and active transformations of bases \cite{MC-BW}.

There is no room for two completely different structures here so quantum information theory {\it must\/} be a sort of 2-spinor calculus. One should not be surprised if certain techniques originating from space-time considerations will find applications in quantum information, or vice versa. Actually, the origin of quantum information processing can be traced back to relativistic considerations since Finkelstein's ``space-time code" \cite{F1} was probably the first paper where the notion of a quantum algorithm occurred.

The goal of this paper is to explore some mutual relations of that kind. I will begin with the celebrated teleportation algorithm, a procedure for exchanging unknown quantum states between different laboratories. I will show that the structure of the algorithm, when translated into the 2-spinor abstract-index language, reveals structures we know from Lorentzian geometry. The observation that metric tensors are linear maps whose formal structures resemble teleportation algorithms may, in principle, have some implications for theories where geometry is quantized. The issue requires further studies.

In the context of quantum algorithms a privileged role is played by Minkowski tetrads (being analogous to entangled states), but null tetrads may also prove important albeit in a different context. The point is that the momentum dependence of SU(2) transformations in momentum space leads to problems with the Lorentz invariance of quantum information \cite{PST}. To understand why it is so consider the case (typical of quantum cryptography) of encoding bits into some sort of linear polarization. The momentum dependent SU(2) transformations rotate each momentum component in a different way and thus typically a state which is linearly polarized in one reference frame becomes depolarized in another one. A practical consequence is that switching between inertial frames makes quantum information channels noisy. The noise is fundamental and of relativistic origin, but for cryptographic purposes may be indistinguishable from the presence of an eavesdropper \cite{MC97}. In order to eliminate the noise one may encode information into eigenstates of relativistic spin operators projected on principal null directions (PNDs) of SL(2,C) transformations \cite{MCMW}. The latter possibility is not as appreciated by quantum information community as it deserves. The problem is that it is difficult to imagine how to realize appropriate null-direction measurements in laboratory. One probably needs new insights, more natural for those who imagine space-time in quantum terms than for quantum opticians thinking in terms of polarizers and interferometers. So the second goal of the paper is to draw attention of the relativistic community to the issue of PND qubits. Had I managed to stimulate some research in that direction, the paper would have served its purpose.

I will start with the original teleportation algorithm. In Section~III I will introduce Penrose's abstract indices, pointing out already at that stage certain analogies with quantum information. In Section~IV I describe a space-time analogue of the teleportation algorithm, and in Section~V a similar 2-spinor algorithm which is more than just an analogy of quantum teleportation. In Section~VI I explain in what sense spin-frames can be regarded as qubits, what is the role of null directions in simplification of the formalism, and why the true qubits have to be defined in yet another way. The latter will be done in Section~VII, where qubits will be represented by $\omega$-spinors.

\section{The original teleportation algorithm}

Let us first recall the original teleportation algorithm  \cite{Tel}.
One begins with three distinguishable systems, labeled $A_1$ (first subsystem in the lab of Alice), $A_2$ (second subsystem in the lab of Alice), and $B$ (the lab of Bob). The goal is to create in the lab of Bob a state that is physically identical to some {\it unknown\/} state,
\be
|\phi,A_1\rangle
&=&
\phi^0|0,A_1\rangle
+
\phi^1|1,A_1\rangle,
\ee
located in the lab of Alice, but perform this only by means of measurements made in the lab of Alice, supplemented by instructions sent by Alice to Bob after each measurement.

Formally what we want to do is to physically implement the map
\be
|\phi,A_1\rangle
&=&
\phi^0|0,A_1\rangle
+
\phi^1|1,A_1\rangle\nonumber\\
&\to&
\phi^0|0,B\rangle
+
\phi^1|1,B\rangle
=
|\phi,B\rangle\label{phi B}
\ee
with unknown $\phi^0$, $\phi^1$. Let us note that in principle there is no problem. Just take the unitary operator
\be
U
&=&
|0,B\rangle\langle 0,A_1|
+
|1,B\rangle\langle 1,A_1|,\\
\ee
and $|\phi,B\rangle=U|\phi,A_1\rangle$. Any quantum dynamics acting in a two-dimensional Hilbert space will do the job, the differences between different $U$s boiling down to different choices of Bob's bases.

What is special about the teleportation algorithm is that propagation of the state from Alice to Bob does not involve unitary quantum evolution.
We assume that Alice and Bob share the same 2-particle state
\be
|\Psi_-,A_2B\rangle
&=&
\frac{1}{\sqrt{2}}
\Big(|0,A_2\rangle |1,B\rangle-|1,A_2\rangle |0,B\rangle\Big)
\ee
so that the initial state is
\be
|\phi,A_1\rangle|\Psi_-,A_2B\rangle.
\ee
Now let us define the {\it Bell basis\/} of Alice
\be
|\Psi_\pm,A_1A_2\rangle
&=&
\frac{1}{\sqrt{2}}
\Big(|0,A_1\rangle |1,A_2\rangle\pm|1,A_1\rangle |0,A_2\rangle\Big),\\
|\Phi_\pm,A_1A_2\rangle
&=&
\frac{1}{\sqrt{2}}
\Big(|0,A_1\rangle |0,A_2\rangle\pm|1,A_1\rangle |1,A_2\rangle\Big).
\ee
and the four unitary operators
\be
U_1 &=& -|0,B\rangle\langle 0,B|
-
|1,B\rangle\langle 1,B|,\\
U_2 &=& -|0,B\rangle\langle 0,B|
+
|1,B\rangle\langle 1,B|,\\
U_3 &=& |0,B\rangle\langle 1,B|
+
|1,B\rangle\langle 0,B|,\\
U_4 &=& -|0,B\rangle\langle 1,B|
+
|1,B\rangle\langle 0,B|,
\ee
acting locally in the lab of Bob. A simple calculation then shows that
\begin{widetext}
\be
|\phi,A_1\rangle|\Psi_-,A_2B\rangle
&=&
\frac{1}{2}
\Big(
|\Psi_-,A_1A_2\rangle U_1|\phi,B\rangle
+
|\Psi_+,A_1A_2\rangle U_2|\phi,B\rangle
+
|\Phi_-,A_1A_2\rangle U_3|\phi,B\rangle
+
|\Phi_+,A_1A_2\rangle U_4|\phi,B\rangle
\Big)\nonumber\\
\ee
\end{widetext}
Now we can formulate the teleportation algorithm. Alice performs in her lab measurements corresponding to projections on the elements of the Bell basis, and on the basis of her measurements sends to Bob one of the four instructions:
\be
{\rm result:\,}|\Psi_-,A_1A_2\rangle  &\to& {\rm instruction:\, do\,\,} U_1^{-1},\nonumber\\
{\rm result:\,}|\Psi_+,A_1A_2\rangle  &\to& {\rm instruction:\, do\,\,} U_2^{-1},\nonumber\\
{\rm result:\,}|\Phi_-,A_1A_2\rangle  &\to& {\rm instruction:\, do\,\,} U_3^{-1},\nonumber\\
{\rm result:\,}|\Phi_+,A_1A_2\rangle  &\to& {\rm instruction:\, do\,\,} U_4^{-1},\nonumber
\ee
Bob receives each of the instructions with equal probability. The characteristic feature distinguishing this algorithm from other quantum algorithms is that the result is always the same: No matter what instruction Bob receives, he produces a particle whose state is $|\phi,B\rangle$. Bob's experimental setup may be regarded as a black box producing pure states $|\phi,B\rangle$ that have the same statistical properties as the states $|\phi,A_1\rangle$ possessed by Alice. Let us note that the procedure works even if Bob does not know which state he produces.

In the next Section I will show that the structure of the teleportation algorithm is well known from Minkowski-space geometry, but we first have to switch from ``Alice" and ``Bob" to abstract indices.

\section{Abstract indices}

The indices ($A_1$, $A_2$, and $B$) we have used above played a role of labels, labeling Hilbert spaces of various parties participating in the teleportation protocol. An analogous (but richer) labeling structure exists also in space-time geometry: These are the abstract indices of Penrose (cf. Chapter 2 in \cite{PR}). Let us recall the basic idea of the construction. Let  $\mathfrak S^\cdot$  denote the module of 2-spinor fields. A labelling system, ${\cal L}=\{A,B,\dots,Z,A_0,B_0,\dots, A_1,\dots\}$, labels canonically isomorphic copies of $\mathfrak S^\cdot$ denoted by $\mathfrak S^{A}$, $\mathfrak S^{B}$, $\mathfrak S^{A_0}$, etc. The indices from $\cal L$ do not take numerical values --- they are just labels.  Light face italic indices are always abstract. However, one sometimes also needs numerical indices, 0 and 1, but then we denote them by upright boldface fonts. Accordingly, the symbol $\phi^A$ denotes a 2-spinor from the copy $\mathfrak S^{A}$ (``a spinor of Alice"), but $\phi^{\bf A}$ may equal $\phi^{0}$ or $\phi^{1}$. $\phi^A$ is basis independent, but $\phi^{\bf A}$ implicitly depends on a basis. The dual of the module $\mathfrak S^{A}$ is denoted by $\mathfrak S_{A}$, and consists of lower-index spinors $\phi_A$. Taking tensor products of a number of 2-spinor modules one arrives at a general module $\mathfrak S^{P \dots R}_{S \dots U}$ of spinors. The isomorphisms between spinors and their duals are denoted by $\ve^{AB}$ and $\ve_{AB}$, and act as follows: $\phi_A=\phi^B\ve_{BA}=\ve_{BA}\phi^B$, $\phi^A=\ve^{AB}\phi_B$. The order of indices is important  since $\ve_{BA}=-\ve_{AB}$, $\ve^{BA}=-\ve^{AB}$. (Thinking in matrix terms, we can say that to lower $A$ we act from right, $\ve_{AB}: \mathfrak S^{A}\to \mathfrak S_{B}$, and to raise it we act from left, $\ve^{AB}: \mathfrak S_{B}\to \mathfrak S^{A}$). The isomorphisms between different copies of 2-spinors of the same type are denoted by $\ve{_A}{^B}$, i.e.
$\ve{_A}{^B}\phi_B=\phi_A$, $\phi^A\ve{_A}{^B}=\phi^B$, $\ve{_A}{^B}: \mathfrak S_{B}\to \mathfrak S_{A}$, $\ve{_A}{^B}: \mathfrak S^{A}\to \mathfrak S^{B}$.
Accordingly, the formula $\phi^A\ve{_A}{^B}=\phi^B$ can be read: ``Shifting Alice's $\phi$ into the space of Bob". The map $\ve{_A}{^B}$ is not yet exactly teleportation, but is very close to it, as we shall see later. Abstract indices not only can be raised or lowered but also (anti)symmetrized and contracted. The basic contraction is $\phi_A\psi^A=-\phi^A\psi_A=\phi_{\bf A}\psi^{\bf A}=\phi_{0}\psi^{0}+\phi_{1}\psi^{1}$
(the summation convention is applied throughout the paper) where the components are taken in arbitrary basis, but the whole expression is basis independent. The rule for raising and lowering numerical indices is $\phi_0=-\phi^1$, $\phi_1=\phi^0$. One needs the operation of complex conjugation, $\overline{\phi^A}=\bar\phi^{A'}\in \mathfrak S^{A'}$. The complex conjugated $\bar\phi^{A'}$ is an entity of a new type, so that an additional set of primed indices is needed, but $\overline{\bar\phi^{A'}}=\phi^A\in \mathfrak S^{A}$. Isomorphisms that map between different copies of $\mathfrak S^{A'}$ and $\mathfrak S_{A'}$ are denoted by $\ve^{A'B'}$, $\ve_{A'B'}$, and $\ve{_{A'}}{^{B'}}$.

Of particular importance for the formalism are the spinors $o^{A}$, $\ii^{A}$, $o^{A'}$, $\ii^{A'}$, known as spin-frames, normalized by $o_{A}\ii^{A}=o_{A'}\ii^{A'}=1$. One can check that they are equivalent to the usual basic qubits and play a role of orthogonal bases in
$\mathfrak S^{A}$ and $\mathfrak S^{A'}$, respectively. The important formula
\be
\ve^{AB}=o^A\ii^B-\ii^A o^B\label{ve}
\ee
is independent of the choice of spin-frames.
Somewhat anticipating our further analysis let us stress here that Eq.~(\ref{ve}) shows that $\ve^{AB}$ is, up to normalization, the entangled state
$|\Psi_-,AB\rangle$ shared by Alice and Bob.

One of the central results of the abstract-index formalism is the identification of the module $\mathfrak S^{AA'}$ (tensor product of primed and unprimed 2-spinor fields) with the one of world-vector fields $\mathfrak S^{a}$. The abstract index $a$ labels world-vector fields, i.e. $x^a\in \mathfrak S^{a}$ is a world-vector (``a position $x$ in the Minkowski space of Alice's configurations..."). The numerical values of the ordinary (non-abstract) upright boldface index $\bf a$ take values 0, 1, 2, and 3. Now, since $\mathfrak S^{a}=\mathfrak S^{AA'}$, we are allowed to write $x^a=x^{AA'}$, although $x^{\bf a}=x^{\bf{AA}'}$ would be meaningless. Instead, we have $x^{\bf a}=g^{\bf a}{_{\bf{AA}'}}x^{\bf{AA}'}$, where $g^{\bf a}{_{\bf{AA}'}}$ denote the so-called Infeld-van der Waerden symbols.
It is convenient for computations that although $x^{\bf a}\neq x^{\bf{AA}'}$, one nevertheless finds
$x^{\bf a}y_{\bf a}= x^{\bf{AA}'}y_{\bf{AA}'}$ for any $x^a$ and $y^a$.
The form of Infeld-van der Waerden symbols varies from basis to basis, but if one takes the Minkowski tetrad $t^a$, $x^a$, $y^a$, $z^a$, defined below, one recognizes (up to normalization) in $g^{0}{_{\bf{AA}'}}$ the $2\times 2$ unit matrix, and $g^{j}{_{\bf{AA}'}}$, $j=1,2,3$ become the three Pauli matrices (cf. Section 3.1 in \cite{PR}).
A Minkowski tetrad, by definition, consists of any four world-vectors satisfying $t_at^a=1$, $x_ax^a=y_ay^a=z_az^a=-1$, with the remaining contractions vanishing. Now consider a spin-frame $o^A$, $\ii^A$, and $o^{A'}=\overline{o^{A}}$, $\ii^{A'}=\overline{\ii^{A}}$ (following \cite{PR} we skip the bars over the primed basis; this does not mean the spin-frames are real, this is just a simplified notation). One checks that the four world-vectors
\be
t^a
&=&
\frac{1}{\sqrt{2}}(o^Ao^{A'}+\ii^A\ii^{A'}),\label{t}\\
x^a
&=&
\frac{1}{\sqrt{2}}(o^A\ii^{A'}+\ii^A o^{A'}),\label{x}\\
y^a
&=&
\frac{i}{\sqrt{2}}(o^A\ii^{A'}-\ii^A o^{A'}),\label{y}\\
z^a
&=&
\frac{1}{\sqrt{2}}(o^Ao^{A'}-\ii^A\ii^{A'}),\label{z}
\ee
define a Minkowski tetrad. Again, anticipating further results, let us note the formal similarity between the above tetrad and the two-qubit Bell basis. (Note that $y^a$ is not proportional to $\ve^{AB}$ since the former involves both primed and unprimed indices.) The metric tensor and the fundamental isomorphism of different copies, i.e. the maps $g^{ab}: \mathfrak S_{b}\to\mathfrak S^{a}$, $g_{ab}: \mathfrak S^{a}\to\mathfrak S_{b}$,
$g{_a}{^{b}}: \mathfrak S^{a}\to\mathfrak S^{b}$, $g{_a}{^{b}}: \mathfrak S_{b}\to\mathfrak S_{a}$,
satisfy
\be
g^{ab}
&=&
t^at^b-x^ax^b-y^ay^b-z^az^b\label{g}\\
&=&
\ve^{AB}\ve^{A'B'}\label{ee}\\
&=&
(o^{A}\ii^{B}-\ii^{A} o^{B})(o^{A'}\ii^{B'}-\ii^{A'} o^{B'})\label{ee'}
\ee
(lowering appropriate indices we obtain $g{_a}{^{b}}$ and $g{_a}{_{b}}$). Eq.~(\ref{g}) is the Minkowski-space resolution of unity. The form (\ref{ee'}) can be used to represent $g^{ab}$ in terms of the null tetrad, i.e. to resolve unity in a null basis,
\be
l^a
&=&
o^Ao^{A'},\label{l}\\
m^a
&=&
o^A\ii^{A'},\label{m}\\
\bar m^a
&=&
\ii^A o^{A'},\label{bar m}\\
n^a
&=&
\ii^A\ii^{A'},\label{n}
\ee
similar in form to the 2-qubit product basis. The metric tensor now reads
\be
g^{ab}
&=&
n^al^b+l^an^b-\bar m^am^b-m^a\bar m^b\label{g nul}.
\ee
All antisymmetric $\phi^{AB}$ are proportional to $\ve^{AB}$. This property, combined with (\ref{ee}), implies that for any world-vector $f_a$ one finds $f_{a}f_{b}\ve^{A'B'}=f_{AA'}f{_{B}}{^{A'}}=\frac{1}{2}f_cf^c \ve_{AB}$.

\section{Space-time analogue of the teleportation algorithm}

So far this has all been the standard textbook material, but let us take a closer look at the following simple calculation
\be
\phi^A\ve^{A'B'}f_{a}f_{b}=\frac{1}{2}f_{c}f^{c}  \phi^A \ve_{AB}=\frac{1}{2}f_{c}f^{c}\phi_B.\label{tele1}
\ee
I claim that this is basically a step of the teleportation algorithm.

We begin with the observation that the part
$\phi^A\ve^{A'B'}$ describes an initial uncorrelated state of a general qubit $\phi^{A}$ (i.e. $|\phi,A\rangle$) of Alice and of the Bell-basis state $\ve^{A'B'}$ (i.e. $|\Psi_-,A'B'\rangle$), where the first bit belongs to Alice, and the second one to Bob.
Let us now decompose the state in a basis, $\phi^{A}=\phi^0 o^{A}+\phi^1\ii^{A}$. Then
\be
\phi^A\ve^{A'B'}t_{a}
&=&
\frac{1}{\sqrt{2}}(\phi^0 \ii^{B'}-\phi^1 o^{B'}),\label{ttt}\\
\phi^A\ve^{A'B'}x_{a}
&=&
\frac{1}{\sqrt{2}}(\phi^0 o^{B'}-\phi^1 \ii^{B'}),\label{xxx}\\
\phi^A\ve^{A'B'}y_{a}
&=&
-\frac{i}{\sqrt{2}}(\phi^0 o^{B'}+\phi^1 \ii^{B'}),\label{yyy}\\
\phi^A\ve^{A'B'}z_{a}
&=&
-\frac{1}{\sqrt{2}}(\phi^0 \ii^{B'}+\phi^1 o^{B'}).\label{zzz}
\ee
The above four spinors are analogous to the states Bob has in his system just after being informed by Alice about her results, but before following her instructions. Only in the case she projected on the analogue of the state $|\Psi_-,AA'\rangle$ (that is, $y_{a}$) the state of Bob's qubit does not require any action. In the remaining cases Bob has to reshuffle the components $\phi^0$ and $\phi^1$, and correct the signs. The actions analogous to $U_j^{-1}$ are
\be
\frac{1}{\sqrt{2}}(\phi^0 \ii^{B'}-\phi^1 o^{B'})t{^{B}}{_{B'}}
&=&
\frac{1}{2}(\phi^0 o^{B}+\phi^1 \ii^{B})
,\label{ttt'}\\
\frac{1}{\sqrt{2}}(\phi^0 o^{B'}-\phi^1 \ii^{B'})x{^{B}}{_{B'}}
&=&
-\frac{1}{2}(\phi^0 o^{B}+\phi^1 \ii^{B})
,\label{xxx'}\\
-\frac{i}{\sqrt{2}}(\phi^0 o^{B'}+\phi^1 \ii^{B'})y{^{B}}{_{B'}}
&=&
-\frac{1}{2}(\phi^0 o^{B}+\phi^1 \ii^{B})
,\label{yyy'}\\
-\frac{1}{\sqrt{2}}(\phi^0 \ii^{B'}+\phi^1 o^{B'})z{^{B}}{_{B'}}
&=&
-
\frac{1}{2}(\phi^0 o^{B}+\phi^1 \ii^{B})
\label{zzz'}
\ee
The right-hand sides of (\ref{ttt'})--(\ref{zzz'}) exhibit the characteristic feature of teleportation algorithms: They are all proportional to the same vector $\phi^{B}$. All these operations occur, via (\ref{g}), in
\be
\phi^A\ve^{A'B'}g{_{a}}{^{B}}{_{B'}}=2\phi^B. \label{2 phi}
\ee
It remains to understand the meaning of the contractions in (\ref{ttt'})--(\ref{zzz'}). We have
\be
t{^{B}}{_{B'}}o^{B'}
=
-\frac{1}{\sqrt{2}}\ii^{B},\quad
t{^{B}}{_{B'}}\ii^{B'}
=
\frac{1}{\sqrt{2}}o^{B},\label{t++}\\
x{^{B}}{_{B'}}o^{B'}
=
-\frac{1}{\sqrt{2}}o^{B},\quad
x{^{B}}{_{B'}}\ii^{B'}
=
\frac{1}{\sqrt{2}}\ii^{B},\label{x++}\\
y{^{B}}{_{B'}}o^{B'}
=
-\frac{i}{\sqrt{2}}o^{B},\quad
y{^{B}}{_{B'}}\ii^{B'}
=
-\frac{i}{\sqrt{2}}\ii^{B},\label{y++}\\
z{^{B}}{_{B'}}o^{B'}
=
\frac{1}{\sqrt{2}}\ii^{B},\quad
z{^{B}}{_{B'}}\ii^{B'}
=
\frac{1}{\sqrt{2}}o^{B}.\label{z++}
\ee
But these are, of course, the transformations Bob employs in the teleportation algorithm. The only difference is that the names given to the instructions and certain multipliers are different from what we are accustomed to in quantum mechanics. But the multipliers have to be different since $y_{a}$ is equivalent to $i|\Psi_-,AA'\rangle$ and, similarly to the link between Infeld-van der Waerden symbols and the Pauli matrices, where the former differ from the latter by the presence of $1/\sqrt{2}$ (cf. Eq.~(3.1.49) in \cite{PR}), we find an appropriate $1/\sqrt{2}$ normalization factor. Summing up, each of the four terms occurring in (\ref{g}) corresponds to one of the four actions required by the teleportation algorithm.

\section{Two-spinor form of the standard teleportation algorithm}

The algorithm described in the previous section required both primed and unprimed spinors, a property absent in the original algorithm. So, in this section we show what has to be modified if Alice and Bob deal with spinors of the same type.
Let us define
\be
t^{A_1A_2}
&=&
\frac{1}{\sqrt{2}}(o^{A_1}o^{A_2}+\ii^{A_1}\ii^{A_2}),\label{tt}\\
x^{A_1A_2}
&=&
\frac{1}{\sqrt{2}}(o^{A_1}\ii^{A_2}+\ii^{A_1}o^{A_2}),\label{xx}\\
y^{A_1A_2}
&=&
\frac{i}{\sqrt{2}}(o^{A_1}\ii^{A_2}-\ii^{A_1} o^{A_2}),\label{yy}\\
z^{A_1A_2}
&=&
\frac{1}{\sqrt{2}}(o^{A_1}o^{A_2}-\ii^{A_1}\ii^{A_2}).\label{zz}
\ee
Now this is the standard Bell basis for ``Alice$_1$" and ``Alice$_2$" [up to $i$ in (\ref{yy})].
Still, one can verify that $t^{A_1A_2}t_{A_1A_2}=1$,
$x^{A_1A_2}x_{A_1A_2}=y^{A_1A_2}y_{A_1A_2}=z^{A_1A_2}z_{A_1A_2}=-1$,
The ``metric tensor"
\be
g^{A_1A_2B_1B_2}
&=&
t^{A_1A_2}t^{B_1B_2}\label{gggg}
\\
&-&
x^{A_1A_2}x^{B_1B_2}-y^{A_1A_2}y^{B_1B_2}-z^{A_1A_2}z^{B_1B_2}
\nonumber\\
&=&
\ve^{A_1B_1}\ve^{A_2B_2}\\
&=&
(o^{A_1}\ii^{B_1}-\ii^{A_1}o^{B_1})(o^{A_2}\ii^{B_2}-\ii^{A_2}o^{B_2})
\ee
decomposes an arbitrary 2-bit spinor $\phi^{AB}$ into its Bell-basis components
\begin{widetext}
\be
\phi^{A_1A_2}
&=&
\phi^{B_1B_2}
g{_{B_1B_2}}{^{A_1A_2}}\\
&=&
\phi^{B_1B_2}t{_{B_1B_2}}
t{^{A_1A_2}}
-
\phi^{B_1B_2}x{_{B_1B_2}}
x{^{A_1A_2}}
-
\phi^{B_1B_2}y{_{B_1B_2}}
y{^{A_1A_2}}
-
\phi^{B_1B_2}z{_{B_1B_2}}
z{^{A_1A_2}}.
\ee
\end{widetext}
Indeed, the contractions $\phi^{B_1B_2}t{_{B_1B_2}}$, $\phi^{B_1B_2}x{_{B_1B_2}}$, $\phi^{B_1B_2}y{_{B_1B_2}}$, $\phi^{B_1B_2}z{_{B_1B_2}}$ are equivalent (up to phase factors) to the scalar products of a general 2-qubit state with, respectively, $|\Phi_+,A_aA_2\rangle$, $|\Psi_+,A_aA_2\rangle$,
$|\Psi_-,A_aA_2\rangle$, and $|\Phi_-,A_aA_2\rangle$.

The formulas are plagued by repeating pairs of similar indices, but we will not risk getting into conflict with the standard spinor formulas if we allow to clump pairs of indices according to $a'=A_1A_2$, and perform calculations by means of the usual 2-spinor tricks (primed world-vector indices are not used in the standard formalism, so there is no risk of confusion). Assuming this, we can write (\ref{gggg}) in the form (\ref{g}) but with the primed indices $a'$, $b'$.

Now, consider the following analogue of (\ref{tele1})
\be
\phi^{A_1}\ve^{A_2B_1}f_{a'}f_{b'}=\frac{1}{2}f_{C_1C_2}f^{C_2C_1} \phi_{B_2}.\label{tele2}
\ee
[valid if $f_{C_1C_2}=\pm f_{C_2C_1}$ which is the case for (\ref{tt})--(\ref{zz})].
I will show that (\ref{tele2}) is precisely the essential step of the teleportation protocol. The whole teleportation protocol consists of four such steps, all of them occurring in
\be
g_{a'b'}
&=&
t_{a'}t_{b'}-x_{a'}x_{b'}-y_{a'}y_{b'}-z_{a'}z_{b'}\label{g'}.
\ee
The concise formula expressing the entire protocol is
\be
\phi^{A_1}\ve^{A_2B_1}g{_{A_1A_2B_1}}{^{B_2}}=\phi{^{B_2}}.
\ee
As before, the part
$\phi^{A_1}\ve^{A_2B_1}$ describes the initial uncorrelated state of a general qubit $\phi^{A_1}$ of Alice and of the entangled state $\ve^{A_2B_1}$, where the first bit belongs to Alice, and the second one to Bob. Let us follow the algorithm in detail. Alice makes measurements and when Bob receives her instructions the state he deals with is one of these:
\be
\phi^{A_1}\ve^{A_2B_1}t_{a'}
&=&
\frac{1}{\sqrt{2}}(\phi^0 \ii^{B_1}-\phi^1 o^{B_1}),\label{t"}\\
\phi^{A_1}\ve^{A_2B_1}x_{a'}
&=&
\frac{1}{\sqrt{2}}(\phi^0 o^{B_1}-\phi^1 \ii^{B_1}),\label{x"}\\
\phi^{A_1}\ve^{A_2B_1}y_{a'}
&=&
-\frac{i}{\sqrt{2}}(\phi^0 o^{B_1}+\phi^1 \ii^{B_1}),\label{y"}\\
\phi^{A_1}\ve^{A_2B_1}z_{a'}
&=&
-\frac{1}{\sqrt{2}}(\phi^0 \ii^{B_1}+\phi^1 o^{B_1}).\label{z"}
\ee
Now Bob behaves according to the instructions:
\be
\frac{1}{\sqrt{2}}(\phi^0 \ii^{B_1}-\phi^1 o^{B_1})t{_{B_1}}{^{B_2}}
&=&
\frac{1}{2}(\phi^0 o^{B_2}+\phi^1 \ii^{B_2})
,
\label{t--}\\
\frac{1}{\sqrt{2}}(\phi^0 o^{B_1}-\phi^1 \ii^{B_1})x{_{B_1}}{^{B_2}}
&=&
-\frac{1}{2}(\phi^0 o^{B_2}+\phi^1 \ii^{B_2}),\label{x--}\\
-\frac{i}{\sqrt{2}}(\phi^0 o^{B_1}+\phi^1 \ii^{B_1})y{_{B_1}}{^{B_2}}
&=&
\frac{1}{2}(\phi^0 o^{B_2}+\phi^1 \ii^{B_2})
,
\label{y--}\\
-\frac{1}{\sqrt{2}}(\phi^0 \ii^{B_1}+\phi^1 o^{B_1})z{_{B_1}}{^{B_2}}
&=&
-\frac{1}{2}(\phi^0 o^{B_2}+\phi^1 \ii^{B_2})
.
\label{z--}
\ee
We again observe the characteristic feature of teleportation algorithms: The right-hand sides of (\ref{t--})--(\ref{z--}) are proportional to the same vector $\phi^{B_2}$.
Following (\ref{g'}), that is subtracting from (\ref{t--}) the sum of (\ref{x--})--(\ref{z--}), we get $\phi^{B_2}$.

\section{Yes-no observables associated with spin-frames}

It is somewhat disappointing that the null representation (\ref{g nul}) will not lead to an algorithmic interpretation of the four terms occurring in $g^{ab}$. The formula (\ref{2 phi}) is, of course, still true, but is not possible to interpret its four parts in terms of a ``make-measurement-and-send-instruction" algorithm. The projections on the null vectors are mathematically equivalent to measurements in the product basis and, repeating the null-tetrad analogues of the steps (\ref{t"})--(\ref{z--}), it is easy to understand why entanglement is necessary for teleportation. On the other hand, it is also clear here that the Minkowski tetrad can be replaced by any four non-null vectors. The latter observation may be, perhaps, of some use for designing alternative teleportation protocols.

In spite of this negative conclusion I will show that null directions are crucial for the link between qubits and spin-frames.

Let ${^*\!}J{^{ab}}{_{X}}{^{Y}}$ and ${^*\!}J{^{ab}}{_{X'}}{^{Y'}}$ denote dualized generators of, respectively, $(1/2,0)$ and $(0,1/2)$ representations of SL(2,C). The Pauli-Lubanski vector corresponding to the two representations is $W{^a}{_{X}}{^{Y}}=P_b{^*\!}J{^{ab}}{_{X}}{^{Y}}$, $W{^a}{_{X'}}{^{Y'}}=P_b{^*\!}J{^{ab}}{_{X'}}{^{Y'}}$. If $V^a$ is a world-vector then the projections $W{_{X}}{^{Y}}=V_aW{^a}{_{X}}{^{Y}}$,
$W{_{X'}}{^{Y'}}=V_aW{^a}{_{X'}}{^{Y'}}$, are ``yes--no" observables whose eigenvectors can play a role of relativistic qubits. The typical choice of timelike $V^a$ leads to the helicity formalism. I will now show that certain null $V^a$ lead to qubits with interesting properties.

If we work in momentum representation then the generator of translations $P^a$ can be identified with
$p^a$, $p_ap^a=m^2$, a future-pointing energy-momentum world-vector. Its decomposition into null components is usually performed in a manner analogous to $t^a$, that is,
\be
p^a &=& \frac{m}{\sqrt{2}}\big(o^A(p)o^{A'}(p)+\ii^A(p)\ii^{A'}(p)\big), \label{p}
\ee
where $o_A(p)\ii^A(p)=1$. Decompositions into null vectors are non-unique, even the number of null directions associated with a given $p^a$ is arbitrary. (The latter freedom is used in the twistor approach to internal symmetries \cite{H} but, in fact, is much more fundamental and allows to introduce SU(N) without any reference to twistors.) In the context of relativistic qubits two null directions are enough, but the problem with (\ref{p}) is the null limit $m\to 0$, essential for optical applications.
From this perspective it is useful to work with spin frames  $\omega_A(p)\pi^A(p)=1$, satisfying
\be
p^a &=& \pi^A(p)\bar\pi^{A'}(p)+(m^2/2)\omega^A(p)\bar\omega^{A'}(p). \label{p'}
\ee
In order to see that the spin frames exist let us take an arbitrary $p$-independent 2-spinor $\nu_A$.
Then
\be
\omega^{A}(p)
&=&
\frac{\nu^A}{
\sqrt{p^{BB'}\nu_B
\bar \nu_{B'}}}
=
\omega^{A}(\nu,p),\label{omega}\\
\pi^A(p)
&=&
\frac{p^{AA'}\bar \nu_{A'}}
{\sqrt{p^{BB'}\nu_B
\bar \nu_{B'}}}
=
\pi^A(\nu, p)\label{pi}
\ee
satisfy (\ref{p'}). If $m\neq 0$, then (\ref{omega}), (\ref{pi}), exist for all $p$; if $m=0$ then there exist $p^a$ parallel to $\nu^a=\nu^A\bar \nu^{A'}$ and (\ref{omega}), (\ref{pi}), have to be defined locally. Treating these spin frames as spinor fields, one finds that
\be
\Lambda\pi_A(\nu, p)&=&\Lambda_{A}{^B}\pi_B(\nu, \Lambda^{-1}p)
=\pi_A(\Lambda\nu, p),\label{pi'}\\
\Lambda\omega_A(\nu, p)&=&\Lambda_{A}{^B}\omega_B(\nu, \Lambda^{-1}p)
=\omega_A(\Lambda\nu, p),\label{omega'}
\ee
meaning that the transformed fields $\Lambda\pi_A(\nu, p)$, $\Lambda\omega_A(\nu, p)$ also satisfy (\ref{p'}).

Taking $V^a=\omega^A(p)\bar\omega^{A'}(p)$ we find
\be
W{_{X}}{^{Y}}(p)
&=&
\frac{1}{2}\big(\pi_X( p)\omega^Y( p)+\omega_X( p)\pi^Y( p)\big),\label{W}\\
W{_{X'}}{^{Y'}}(p)
&=&
-\frac{1}{2}\big(\bar\pi_{X'}( p)\bar\omega^{Y'}( p)+\bar\omega_{X'}( p)\bar\pi^{Y'}( p)\big).\label{W'}
\ee
The spin-frame simultaneously solves the eigenvalue problem
\be
W{_{X}}{^{Y}}(p)\omega_Y( p)
&=&
\frac{1}{2}\omega_X( p),\label{W1+}\\
W{_{X'}}{^{Y'}}(p)\bar\pi_{Y'}( p)
&=&
\frac{1}{2}\bar\pi_{X'}( p),\label{W2+}\\
W{_{X}}{^{Y}}(p)\pi_Y( p)
&=&
-\frac{1}{2}\pi_X( p),\label{W2-}\\
W{_{X'}}{^{Y'}}(p)\bar\omega_{Y'}( p)
&=&
-\frac{1}{2}\bar\omega_{X'}( p),\label{W1-}
\ee
which explains why in our discussion of teleportation the identification of spin-frames with qubits was more than just an analogy.
The simplicity of (\ref{W})--(\ref{W1-}) would be lost if one took a non-null $V^a$, unless $m=0$ where all choices of $V^a$ are equivalent (because then $W^a$ and $P^a$ are parallel, and different components of $W^a$ commute).

One should bear in mind that the above spin-frames cannot be regarded as qubits in the standard quantum-mechanical meaning of this word. Qubits belong to carrier spaces of unitary representations, whereas the spin-frames carry non unitary representations of SL(2,C).
In the next section we shall show how to associate with $\omega_X( p)$, $\pi_X( p)$, the true qubits, i.e. spinor fields transforming unitarily. We will deal with a new type of indices, but the whole 2-spinor teleportation framework will remain valid. The construction of the representation will be performed in the 2-spinor formalism. Instead of inducing from little groups I will use here the duality between active finite-dimensional non-unitary SL(2,C) transformations of spin-frames, and passive unitary infinite-dimensional transformations of Bargmann-Wigner amplitudes \cite{MC-BW}.

\section{True relativistic qubits}

Let $d\tilde p=d^3p/[(2\pi)^32p_0]$, $px=p_ax^a$, and consider a positive-energy solution of the Dirac equation, expressed in the basis of eigenvectors of the Pauli-Lubanski vector
\be
\left(
\begin{array}{cc}
W{_{X}}{^{Y}}(p) & 0\\
0 & W{_{X'}}{^{Y'}}(p)
\end{array}
\right)
\ee
of the $(1/2,0)\oplus(0,1/2)$ representation, i.e.
\be
\left(
\begin{array}{c}
\phi_A(x)\\
\phi_{A'}(x)
\end{array}
\right)
&=&\int d\tilde p\left[
\left(
\begin{array}{c}
-\pi_A(p)\\
-\frac{m}{\sqrt{2}}\bar\omega_{A'}(p)
\end{array}
\right)
\phi_{\it 0}(p)
+
\left(
\begin{array}{c}
\frac{m}{\sqrt{2}}\omega_{A}(p)\\
-\bar\pi_{A'}(p)
\end{array}
\right)
\phi_{\it 1}(p)\right]e^{-ipx}\label{Dir}
\ee
The amplitudes $\phi_{\it 0}(p)$, $\phi_{\it 1}(p)$ involve italic numerical indices to distinguish them from the ordinary SL(2,C) 2-spinors.
Since $d\tilde p$ is Poincar\'e invariant, we find
\be
\left(
\begin{array}{c}
\Lambda{_{A}}{^{B}}\phi_B(\Lambda^{-1}x)\\
\Lambda{_{A'}}{^{B'}}\phi_{B'}(\Lambda^{-1}x)
\end{array}
\right)
&=&\int d\tilde p\left[
\left(
\begin{array}{c}
-\Lambda\pi_A(p)\\
-\frac{m}{\sqrt{2}}\overline{\Lambda\omega}{_{A'}}(p)
\end{array}
\right)
\phi_{\it 0}(\Lambda^{-1}p)
+
\left(
\begin{array}{c}
\frac{m}{\sqrt{2}}\Lambda\omega_{A}(p)\\
-\overline{\Lambda\pi}{_{A'}}(p)
\end{array}
\right)
\phi_{\it 1}(\Lambda^{-1}p)\right]e^{-ipx}\label{Dir1}\\
&=&\int d\tilde p\left[
\left(
\begin{array}{c}
-\pi_A(p)\\
-\frac{m}{\sqrt{2}}\bar\omega_{A'}(p)
\end{array}
\right)
\Lambda\phi_{\it 0}(p)
+
\left(
\begin{array}{c}
\frac{m}{\sqrt{2}}\omega_{A}(p)\nonumber\\
-\bar\pi_{A'}(p)
\end{array}
\right)
\Lambda\phi_{\it 1}(p)\right]e^{-ipx},\label{Dir2}
\ee
where
\be
\Lambda\omega_A(p) &=& \Lambda_{A}{^B}\omega_B(\Lambda^{-1}p),\\
\Lambda\pi_A(p) &=& \Lambda_{A}{^B}\pi_B(\Lambda^{-1}p),
\ee
and $\Lambda^{-1}p$ stands for $\Lambda^{-1}{_{A}}{^{B}}\bar\Lambda^{-1}{_{A'}}{^{B'}}p_b$.
(\ref{Dir2}) defines implicitly the new amplitudes $\Lambda\phi^{\it 0}(p)$, $\Lambda\phi^{\it 1}(p)$. We immediately find that
\be
\left(
\begin{array}{c}
\Lambda\phi_{\it 0}(p)\\
\Lambda\phi_{\it 1}(p)
\end{array}
\right)
&=&
\left(
\begin{array}{cc}
\omega_{A}(p)\Lambda\pi^{A}(p) & -\frac{m}{\sqrt{2}}\omega_{A}(p)\Lambda\omega^{A}(p)\\
\frac{m}{\sqrt{2}}\bar\omega_{A'}(p)\overline{\Lambda\omega}{^{A'}}(p) & \bar\omega_{A'}(p)\overline{\Lambda\pi}{^{A'}}(p)
\end{array}
\right)
\left(
\begin{array}{c}
\phi_{\it 0}(\Lambda^{-1}p)\\
\phi_{\it 1}(\Lambda^{-1}p)
\end{array}
\right)\label{true}.
\ee
The matrix in (\ref{true}) belongs to SU(2) and will be denoted by $\Lambda(p){_{\pmb A}}{^{\pmb B}}$ (italic boldface indices; the corresponding abstract indices will be denoted by ${\cal A}$, ${\cal B}$, i.e. by the calligraphic font).
Writing Eq.~(\ref{true}) as
\be
\Lambda\phi_{\cal A}(p)
&=&
\Lambda(p){_{\cal A}}{^{\cal B}}
\phi_{\cal B}(\Lambda^{-1}p)\label{true'}
\ee
we obtain a compact and manifestly covariant form of the unitary representation of the (covering of) the Poincar\'e group whose spin is 1/2 and mass is $m$. (Four-translations by $y$ reduce to the usual multiplication by $e^{ipy}$ --- we skip this point). Spinor-fields $\phi_{\cal A}(p)$ are sometimes termed the $\omega$-spinors or Bargmann-Wigner 2-spinors, and $\Lambda(p){_{\cal A}}{^{\cal B}}$ is the Wigner rotation. If $m=0$ the matrix is diagonal and the diagonal elements, $\Lambda(p){_{\it 0}}{^{\it 0}}$, $\Lambda(p){_{\it 1}}{^{\it 1}}$, are phase factors, a consequence of the fact that both
$\Lambda\pi_{A}(p)$ and $\pi_{A}(p)$ have the same flagpole
\be
p_a=\pi_{A}(p)\bar\pi_{A'}(p)=\Lambda\pi_{A}(p)\overline{\Lambda\pi}_{A'}(p).
\ee
Computing generators of (\ref{true'}) we can find the projection of the corresponding Pauli-Lubanski vector on $V^a=\omega^A(p)\bar\omega^{A'}(p)$. The result is simply
\be
W{_{\pmb A}}{^{\pmb B}}(p)
&=&
-\frac{1}{2}
\left(
\begin{array}{cc}
1 & 0\\
0 & -1
\end{array}
\right)
\ee
The fact that $\Lambda(p){_{\cal A}}{^{\cal B}}$  is $p$-dependent has consequences for quantum information processing. For example, in optics states of linear polarization are linear combinations of ``timelike" circular polarizations (i.e. eigenstates of the helicity operators, $W^0{_A}{^B}(p)$, etc.). Since different momentum components are multiplied by different, momentum dependent Wigner phase factors, each Fourier component of a linearly polarized light gets rotated by a different Wigner angle. As a result, SL(2,C) transformations map linear polarization states into depolarized states, and thus introduce into relativistic communication systems a kind of noise, whose origin is purely relativistic. Entropy computed on the basis of a ``spin" reduced density matrix (obtained by tracing out momenta from the full momentum-dependent density matrix of the qubit) provides a quantitative measure of the effect. This observation started an ongoing discussion on the meaning of relativistic effects for quantum information processing.

However, if one works with {\it null\/} circular polarizations, that is the eigenstates of the Pauli-Lubanski vector projected on $\omega^A(p)\bar\omega^{A'}(p)$, the situation changes. In order to understand it, let us return to $\omega^A(\nu,p)$, which is parallel to an arbitrarily chosen and momentum independent 2-spinor $\nu^A$. It is clear that eigenvectors of the projection of the Pauli-Lubanski vector will not change if one replaces $\omega^A(p)\bar\omega^{A'}(p)$ by $\nu^A\bar\nu^{A'}$. This remark is important sice projecting on $p$-dependent directions is even less clear than projecting on directions that are null.

Choosing $\nu^A$ satisfying $\Lambda{_{A}}{^{B}}\nu_B=\lambda\, \nu_A$ we find
\be
\omega_{A}(p)\Lambda\pi^{A}(p) &=& \bar \lambda/|\lambda|=e^{-i\phi}
\ee
and
\be
\Lambda{_{\pmb A}}{^{\pmb B}}(p)
&=&
\left(
\begin{array}{cc}
e^{-i\phi} & 0\\
0 & e^{i\phi}
\end{array}
\right)
\ee
where $\phi$ does not depend on $p$. Let us stress that $m=0$ was not assumed. So replacing the helicity by an appropriately adjusted null circular polarization removes the depolarization effect: All wavelengths undergo rotation of polarization by the same angle $\phi$. The entropy of an appropriate ``spin" reduced density matrix will be independent of $\Lambda$. In effect, the procedure of adjusting qubits to Lorentz transformation plays a role of relativistic error correction.

\section{Summary}

Tensor products of 2-spinors have geometric meaning. Tensor products of qubits are interpretable in terms of information. Labels employed in quantum information (Alice, Bob, Eve...) have the same status as the Penrose abstract indices. The antisymmetric two-index spinor $\ve^{AB}$ is the same object as the maximally entangled Einstein-Podolsky-Rosen state shared by Alice and Bob. It plays a fundamental role for both space-time considerations and quantum information processing, but its standard applications in the two domains are completely different and apparently unrelated.
Pursuing further this analogy, if we take the tensor product $\ve^{AB}\ve^{A'B'}$ we obtain Alice connected by a quantum channel with Bob, and similarly Alice' connected with Bob', but the two groups working independently of each another. However, geometrically these four parties define a Lorentzian metric in some abstract space-time.
Decomposing this metric into Minkowski tetrads we obtain four geometric objects that formally represent the four steps of the teleportation protocol. The Minkowski tetrads themselves have a form of the Bell basis, another concept crucial for quantum communication and information processing.

These facts may be just curiosities, but equally well could be a departure point for some new type of studies on the borderland between classical space-time and quantum physics. One problem that already awaits solution is to physically implement qubits associated with projections of spin on null directions. Qubits associated with PNDs of SL(2,C) transformations might play a fundamental role in noise reduction if Alice and Bob operate in different reference frames. These objects are completely counterintuitive for quantum opticians accustomed to polarizers and interferometers. To implement the new qubits in practice one needs new ideas --- partly geometric and partly quantum.

\end{document}